\documentclass[usenatbib,useAMS,twocolumn]{mn2e}
\usepackage[latin1]{inputenc}
\usepackage{psfig}

\usepackage{epsfig}
\usepackage{color}

\input{psfig.sty}

\def\lsim{\mathrel{\rlap{\lower 4pt \hbox{\hskip 1pt $\sim$}}\raise 1pt\hbox {$<$}}}
\def\gsim{\mathrel{\rlap{\lower 4pt \hbox{\hskip 1pt $\sim$}}\raise 1pt\hbox {$>$}}}

\title{Cosmological Implications of the Second Parameter of Type Ia Supernovae}

\author[Podsiadlowski et al.]
{Philipp Podsiadlowski$^{1}$\thanks{E-mail: podsi@astro.ox.ac.uk},
Paolo A. Mazzali$^{2,3}$, Pierre Lesaffre$^{1,4,5}$, Christian Wolf$^{1}$
\newauthor and Francisco F\"orster$^{1}$\\
$^{1}${\it Dept.\ of Physics, Oxford University, Oxford, OX1 3RH, UK}\\
$^{2}${\it Max-Planck Institut f\"ur Astrophysik,
  Karl-Schwarzschildstr. 1, 85748 Garching, Germany}\\
$^{3}${\it Istituto Naz.\ di Astrofisica-Oss.\ Astron., Via Tiepolo, 11,
  34131 Trieste, Italy}\\
$^{4}${\it Institute of Astronomy, Cambridge CB3 0HA, UK}\\
$^{5}${\it  LRA, 24 rue Lhomond, 75231 PARIS Cedex 05, France}
}
\date{\today}

\volume{000}

\pagerange{1\,--\,10} \pubyear{2006}

\begin{document}
\maketitle

\label{firstpage}

\begin{abstract}
Theoretical models predict that the initial metallicity of the
progenitor of a Type Ia supernova (SN Ia) affects the peak of the
supernova light curve. This can cause a deviation from the standard
light curve calibration employed when using SNe Ia as standardizable
distance candles and, if there is a systematic evolution of the
metallicity of SN Ia progenitors, could affect the determination
of cosmological parameters. Here we show that this metallicity effect
can be substantially larger than has been estimated previously, when
the neutronisation in the immediate pre-explosion phase in the CO
white dwarf is taken into account, and quantitatively assess the
importance of metallicity evolution for determining cosmological
parameters.  We show that, in principle, a moderate and plausible
amount of metallicity evolution could {\em mimic} a
$\Lambda$-dominated, flat Universe in an open, $\Lambda$-free
Universe. However, the effect of metallicity evolution appears not
large enough to explain the high-z SN Ia data in a flat Universe, for
which there is strong independent evidence, without a cosmological
constant.  We also estimate the systematic uncertainties introduced by
metallicity evolution in a $\Lambda$-dominated, flat Universe. We find
that metallicity evolution may limit the precision with which
$\Omega_{\rm m}$ and $w$ can be measured and that it will be difficult
to distinguish evolution of the equation of state of dark energy from
metallicity evolution, at least from SN Ia data alone.
\end{abstract}


%%%%%%%%%%%%%%%%%%%%%%%%%%%%%%%%%%%%%%%%%%%%%%%%%%%%%%%%%%%%%%%%%%%%%%%%%%%%%%

\section{Introduction}

The use of Type Ia supernovae (SNe Ia) as standardizable cosmological
distance candles (Riess et al.\ 1998; Perlmutter et al.\ 1999; Tonry
et al.\ 2003; Riess et al.\ 2004; Astier et al.\ 2006) relies on the
empirical fact that there is a tight correlation between the supernova
peak brightness and the width of the supernova light curve (Phillips
1993), i.e. the fact that, to lowest order, SN Ia light curves form a
one-parameter family of curves. 
The driving parameter that determines the relation, assuming that the mass of
the progenitor white dwarf at explosion is constant, has been shown to be the
opacity in the ejecta (Khokhlov, M\"uller \& H\"oflich 1993;
H\"oflich et al.\ 1996, Mazzali et al.\ 2001). This is closely related to the
quantity of radioactive $^{56}$Ni synthesised in the explosion, which is
responsible for the SN luminosity.  
In recent years it has become apparent
from a larger sample of observed supernovae that not all SN Ia light
curves fit into this one-parameter family, producing an intrinsic,
scatter in the Phillips relation (see the discussion in Mazzali \&
Podsiadlowski [2006] and Benetti et al.\ [2004]).  This immediately
implies that there must be more than one parameter controlling SN Ia
light curves. The physical property in the progenitor that determines
the dominating (first) parameter still has not been clearly
identified. On the other hand, from a theoretical point of view it
seems unavoidable that the metallicity of the original supernova
progenitor must at least in part be responsible for the intrinsic
scatter about the mean Phillips relation.  Timmes, Brown \& Truran
(2003) showed, using straightforward and uncontroversial nuclear
physics arguments, that the neutron excess in the immediate progenitor
white dwarf, which is a direct function of the initial metallicity of
the progenitor star, controls the fraction of radioactive to
non-radioactive Ni produced in the exploding white dwarf and hence
affects the peak supernova luminosity, an effect subsequently
confirmed by detailed explosion calculations (R\"opke et al.\ 2005;
Travaglio, Hillebrandt \& Reinecke 2005). On the other hand, the width
of the supernova light curve is ultimately determined by the opacity
of the ejecta which is mainly a function of the total amount of
iron-peak elements, i.e. is independent of the ratio of radioactive to
non-radioactive material (Mazzali et al.\ 2001).  Mazzali \&
Podsiadlowski (2006) demonstrated that this produces an intrinsic
scatter around the Phillips relation comparable to the observed
scatter. An intrinsic scatter in the Phillips relation in itself does
not necessarily limit the usefulness of SNe Ia as cosmological probes,
as long as the scatter is non-systematic.  However, a physical
parameter such as metallicity can in principle introduce subtle
evolutionary effects that could affect the determination of
cosmological parameters. It is the purpose of this paper to quantify
the magnitude of such evolutionary effects and to demonstrate that
these need to be taken into account, in particular when trying to
constrain higher-order effects such as the cosmological evolution of
the equation of state of dark energy or its variants.

In Section~2 we show that the effect of metallicity on the supernova
light curves may actually be significantly larger than originally
estimated by Timmes et al.\ (2003), since these authors did not
consider the additional neutronisation in the white dwarf core in the
immediate pre-explosion phase. In Section~3 we investigate
quantitatively the conditions under which the apparently observed
deviation in the Hubble relation from a $\Lambda$-free cosmology could
be explained by evolutionary effects, and in Section~4 we estimate the
metallicity-dependent corrections when determining the equation of state
of dark energy in a $\Lambda$-dominated cosmology, as presently
favoured by various experiments. Finally in Section 5 we discuss
these results and possible observational tests.

\section{Relation between \boldmath{$M_{\rm peak}$} and \boldmath{$Z$}}

As Timmes et al.\ (2003) showed, the neutron excess in the core of a
CO white dwarf at the time of the thermonuclear runaway determines the
ratio of stable to unstable Ni, which in turn determines the supernova
peak luminosity.  The neutron excess depends on the abundance of
elements that have an excess of neutrons, such $^{56}$Fe and
$^{22}$Ne.  The main source of excess neutrons in the interstellar
medium from which a star is born is $^{56}$Fe which has four extra
neutron in each nucleus. The neutron excess\footnote{The neutron
excess, $\eta_i$, for a nucleus of species $i$ can be written as
$\Delta n\,Y_i$, where $\Delta n$ is the number of extra neutrons per
nucleus (i.e. the number of neutrons minus the number of protons) and
$Y_i$ is the number fraction for this species. The total $\eta$,
summed over all species, is directly related to the electron number
fraction, $Y_e$, according to $\eta = 1 - 2Y_e$.}
for $^{56}$Fe therefore is
\begin{equation}
\eta_{56}=4\, X(^{56}{\rm Fe})/56,
\end{equation}
where we use the notation $X({\rm E})$ to denote the mass fraction of
a given element E. The abundance of $^{56}$Fe is not modified until
the explosion (any gravitational settling will be removed by the
growth of the convective core during the carbon flash [C-flash]).

In contrast, $^{22}$Ne is formed predominantly during the progenitor's
nuclear evolution: during hydrogen burning, the CNO cycle converts
most of the initial nuclei of $^{12}$C, $^{16}$O into $^{14}$N, which
subsequently is converted into $^{22}$Ne by two $\alpha$
captures during helium burning. Thus, essentially all of the initial
abundances of C, N and O are converted into $^{22}$Ne in the
white-dwarf core, making it the most abundant element after $^{12}$C
and $^{16}$O\footnote{Note that the $^{12}$C and $^{16}$O nuclei that
exist at the time of the explosion have been exclusively produced in
the progenitor star by the triple-$\alpha$ reaction and $\alpha$
captures during helium burning.}. Hence there is a direct relation
between the $^{22}$Ne contained in the C+O WD and the amount of C, N
and O nuclei initially present in the star:
\begin{equation}
\frac1{22}X(^{22}{\rm Ne})=\frac1{12}X(^{12}{\rm C})+
\frac1{14}X(^{14}{\rm N})+
\frac1{16}X(^{16}{\rm O}).
\end{equation}
Since $^{16}$O is usually the most abundant element by number in
star-forming regions, it is the initial $^{16}$O abundance that is the
most important element for determining the neutron excess in the
white-dwarf core.

Timmes et al.\ (2003) assumed that the neutron excess at the time of
the explosion was entirely determined by the initial abundances of C,
N, O and Fe. However, before carbon burning runs away and initiates
the supernova explosion, there is a long phase of low-level carbon
burning, typically lasting several thousand years, during which the
excess energy produced by carbon burning is efficiently transported
away from the burning region by convection, delaying the thermonuclear
runaway and leading to a gradually increasing convective core. In this
phase, the densities are high enough that electron captures are very
efficient and further neutronise the core material.  To model this
realistically, one needs detailed stellar evolution calculations with
a detailed nuclear reaction work and a proper treatment of the
convective Urca process (see Lesaffre, Podsiadlowski \& Tout 2005).
However, with some approximations, we can estimate the effect of
additional electron captures onto $^{22}$Ne prior to the explosion
phase. Each electron capture (or inverse beta decay) increases the
number of excess neutrons per nucleus by two (since one proton is
converted into one neutron). On the other hand, at these high
densities the inverse reactions (beta decays, positron captures) are
strongly disfavoured energetically. Therefore, the change in the
overall neutronisation of the matter depends only on the total number
of electron captures/inverse beta decays.

For this analysis, we consider the reduced network of the main
nuclear reactions during carbon burning shown in table 6 of Arnett \&
Thielemann (1985).  In this network, we further neglect
neutron-capture reactions compared to $\alpha$ and proton captures.
This is reasonable since the main basic reactions are
$^{12}$C$(^{12}$C$,\alpha)^{20}$Ne and $^{12}$C$(^{12}$C$,p)^{23}$Na
which produce $\alpha$ particles and protons in approximately equal
amounts, whereas neutron sources come from much weaker reactions.

We now consider the fate of $^{22}$Ne in this network and record the
net number of electron captures involved in the subsequent chain of
reactions.  Each electron capture will indeed add two extra
neutrons/nucleus. Neglecting the neutron capture reaction, $^{22}$Ne
has one of two choices:
\begin{itemize}
  \item it can either capture an $\alpha$ particle and be locked
in $^{26}$Mg via the chain $^{22}$Ne$(\alpha,n)^{25}$Mg$(n,\gamma)^{26}$Mg,
  \item or it can capture a proton via  $^{22}$Ne$(p,\gamma)^{23}$Na. 
\end{itemize}
Using analytical rates from the REACLIB (Rauscher \& Thielemann 2002)
data base, we observe that the branching ratio of these two reactions
is more than 97\% in favour of the proton capture for all temperatures
below $2\times 10^9$\,K, a temperature which is well below the maximum
temperature that can be expected during the C-flash (Lesaffre et al.\
2006).  We can hence safely neglect the alpha-capture reaction.
  
The $^{23}$Na nucleus produced by the proton capture can now either
capture an electron with a probability $q$ to become $^{23}$Ne or
capture a proton with a probability $1-q$.  In the latter case, the
chain of reactions is
$^{23}$Na$(p,\gamma)^{24}$Mg$(p,\gamma)^{25}$Al$(\beta^{+})^{25}$Mg
with one inverse beta decay. $^{25}$Mg can capture either an electron
or a proton, but in both cases it leads to one electron
capture/inverse beta decay since
$^{25}$Mg$(p,\gamma)^{26}$Al$(\beta^{+})^{26}$Mg.  The probability $q$
is not easy to estimate since it depends on how the proton-capture
rate compares to the convective Urca process on $^{25}$Mg.
 
A $^{22}$Ne nucleus can hence either capture one electron with a
probability $q$ or capture two electrons with a probability $(1-q)$.
Finally, if we assume that a fraction $f$ of the $^{22}$Ne is burnt
during the C-flash, this leads to an increase of the neutron excess of
$2\,f\,(2-q)$ for each initial $^{22}$Ne nucleus which already has two
extra neutrons. Summing up these contributions, we obtain the total
neutron excess before the C-flash due to $^{22}$Ne 
\begin{equation}
\eta_{22}=2\,[1+f\,(2-q)]\,X(^{22}{\rm Ne})/22. 
\end{equation}
Based on our present C-flash calculations (Lesaffre et al.\ 2006), we
estimate that at least 0.02 of the mass of carbon is burnt in the C-flash
phase, implying that most of the $^{22}$Ne may be consumed (i.e.
$f\simeq 1$). However, this will generally depend on the initial abundances
and also the C/O ratio and needs to be confirmed with a full reaction
network (both $f$ and $q$ could in fact depend on the first parameter!).

Summing up all contributions of metallicity to the neutronisation from
the birth of the star until the start of the explosion, we get:
$$
\eta_Z=\eta_{22}+\eta_{56}= 4\,\frac{X(^{56}{\rm Fe})}{56}
$$
\begin{equation}
\hspace{0.5cm}+2\,[1+f\,(2-q)]\left[\frac{X(^{12}{\rm C})}{12}+
\frac{X(^{14}{\rm N})}{14}+
\frac{X(^{16}{\rm O})}{16}\right] \mbox{.}
\end{equation}

Finally, the neutron excess might change according to parameters other
than the metallicity, and we write the net neutronisation at the time
of the explosion as
\begin{equation}
\eta=\eta_0+\eta_Z,
\end{equation}
where the subscript 0 denotes the neutronisation for zero
metallicity.

Following Timmes et al.\ (2003), we assume that the combustion is fast
enough compared to weak interactions so that the neutron excess does
not change in the process\footnote{This will generally not be correct
for the central region of the white dwarf ($\sim 0.1- 0.2\,M_\odot$)
where the main nucleosynthesis products are $^{54}$Fe, $^{56}$Fe and
$^{58}$Ni (e.g. Nomoto et al.\ 1984). However, this radioactively
inert region is not expected to contribute significantly to the
lightcurve shape (e.g. Mazzali \& Podsiadlowski 2006).}. Allowing only
$^{56}$Ni and $^{58}$Ni as products of the burning, we get the mass
fraction of unstable $^{56}$Ni
\begin{equation}
X(^{56}\mbox{Ni})=1-29\eta \mbox{.}
\end{equation}

Since the peak luminosity $L$ of a SN Ia is roughly proportional to
$X(^{56}$Ni$)$ for a given value of the primary parameter
(e.g. Arnett 1982), we can write
\begin{equation}
L=l_0\, X(^{56}\mbox{Ni}) \mbox{.}
\end{equation}
From this we compute the magnitude difference due to metallicity
effects
\begin{equation}
M-M_0=-2.5 \log_{10}\left(1-\frac{29}{1-29\eta_0}\eta_Z\right),
\end{equation}
where $M$ and $M_0$ are the peak magnitudes for the light curves of
SNe Ia of a given primary parameter with ($M$) and without ($M_0$)
metals.

Setting $\eta_0=0$  (i.e. assuming that the primary parameter does
not affect the neutronisation) and adopting the solar abundance ratios 
from Asplund, Grevesse \& Sauval (2005), we can write equation~(6)
as
\begin{equation}
X(^{56}\mbox{Ni})=1-\alpha\,Z/Z_0 \mbox{.}
\end{equation}
For $Z_0=0.02$, we obtain $\alpha = 0.165$ (for $q=0$ and $f=1$) and
$\alpha = 0.111$ (for $q=1$ and $f=1$)\footnote{The metallicity
normalisation was chosen as $Z_0=0.02$ since this is a typical value
adopted in the literature to represent ``solar metallicity''. With the
Asplund et al.\ (2005) composition, this implies a logarithmic oxygen
abundance by number of $\log {\rm (O/H) + 12} = 8.87$.  In contrast,
the corresponding values for the Sun are 0.0122 and 8.66 for the
Asplund et al.\ (2005) composition mixture, which has a hydrogen mass
fraction $X = 0.7329$, and abundance ratios (by number)  C/O$ =
0.537$, N/O$ = 0.132$ and Fe/O$ = 0.062$.}.  For comparison, with
$f=0$ (i.e.  neglecting the neutronisation during the C-flash), we
recover $\alpha= 0.058$, the value obtained by Timmes et al.\ (2003).

The metallicity dependence of the supernova peak magnitude relative
to a reference model $M_{\rm ref}$ can then be written as
\begin{equation}
M-M_{\rm ref}=-2.5  \log_{10}\left[(1-\alpha Z/Z_0) 
/(1-\alpha Z_{\rm ref}/Z_0) \right],
\end{equation}
where $Z_{\rm ref}$ could, e.g., represent
the typical metallicity of the SN Ia sample used in the calibration
of the Phillips relation. Figure~1 shows this relation for different
values of $\alpha$ for zero metallicity.  Expanding this relation to second
order in $\alpha$, we can approximate this relation as
\begin{equation}
M-M_{\rm ref}\simeq 1.086\alpha\,{Z-Z_{\rm ref}\over Z_0}\,
\left[1 + {1\over 2}\alpha\,{Z+Z_{\rm ref}\over Z_0}\right].
\end{equation}
Since the factor in square brackets is not much larger than 1, this
shows that, to lowest order, the metallicity-dependent correction is
proportional to both $\alpha$ and $(Z-Z_{\rm ref})$. In the following
sections, we generally adopt an intermediate value for $\alpha$ of
0.111, but because of this linear relationship, it is easy to
rescale all the metallicity effects for different values of $\alpha$.

\begin{figure}
\centerline{ \psfig{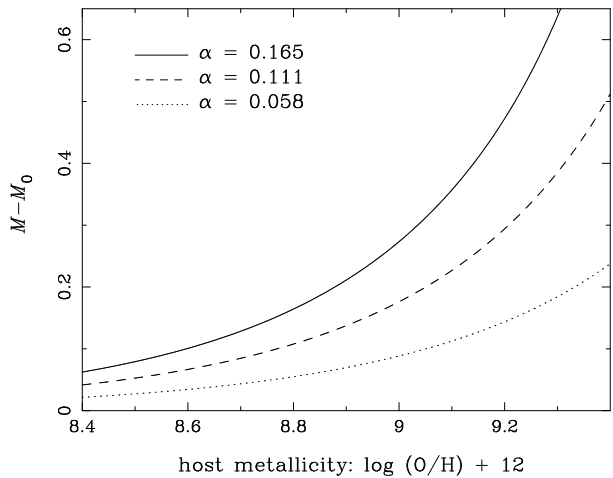} }
\caption{The effect of metallicity on the peak magnitude of SNe Ia,
relative to a reference peak magnitude $M_0$ for $Z=0$, as a
function of logarithmic oxygen abundance (by number). On this scale, a
metallicity of $Z=0.02$ corresponds to $\log \rm(O/H) + 12 = 8.87$
(Asplund et al.\ 2005).}
\label{mz}
\end{figure}

Hamuy et al.\ (2000) and Gallagher et al.\ (2005) have previously
looked for a correlation between SN Ia peak magnitudes and the
metallicity of the host galaxy for nearby SNe Ia and found no
significant correlation. This indeed confirms that metallicity cannot
be the dominant first parameter that controls the variation of SN Ia
light curves but can only be a second parameter. These authors 
suggested that the first parameter is related to age, which in turn
could be related to the central ignition density (see, e.g., Lesaffre et
al.\ 2006).  Gallagher et al.\ (2005) also attempted to identify any
systematic metallicity-dependent deviation from the mean Hubble
relation and found a weak metallicity trend, but one that was not very
statistically significant (also see Wang et al.\ 2006).

Finally, we note that metallicity can modify the properties of the
exploding white dwarfs in other ways than just through the neutron
excess; for the single-degenerate model, various theoretical studies
have predicted or suggested that metallicity affects the initial mass
range for white dwarfs that will ultimately explode (Langer et al.\
2000), the C/O ratio in the white dwarf (H\"oflich, Wheeler \&
Thielemann 1998; Umeda et al.\ 1999) and the accretion efficiency in
the progenitor phase (Kobayashi et al.\ 1998; Umeda et al.\ 1999). All
of these factors could either enhance or reduce the effect of the
neutron excess on its own (the possible evolutionary consequences of
some of these effects have recently also been considered by
Riess \& Livio [2006]).

\section{The effects of metallicity on the determination of cosmological
parameters}

As is clear from Figure~1, a systematic variation in metallicity could
in principle introduce a systematic effect on the observed
SN Ia relation that is comparable in magnitude to the inferred effect
that led to the first evidence for an accelerating Universe
(Riess et al.\ 1998; Perlmutter et al.\ 1999). In this section
we will therefore ask the question what metallicity evolution is required
to {\em mimic} a $\Lambda$-dominated Universe in a more traditional 
$\Lambda$-free cosmology.

\subsection*{\em The Phillips relation and the first and second parameter}

\begin{figure}
\centerline{
\psfig{file=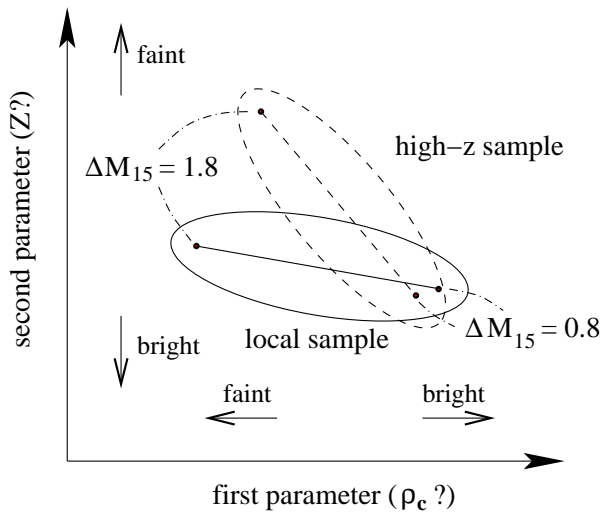,width=8cm}}
\caption{Schematic diagram illustrating possible evolutionary effects
when sampling different local and high-z supernovae assuming two
underlying parameters. }
\end{figure}

Before investigating metallicity effects, it is important to point out
that, while the Phillips relation used to correct SN Ia light curves
is a one-parameter relation, whether this is done using the $\Delta
M_{15}$ method (Phillips et al.\ 1999) or an equivalent method such as
the stretch method (Goldhaber et al.\ 2001), this parameter is
unlikely to be completely independent of metallicity. The calibration
of the relation uses a local sample of SNe Ia which includes
supernovae over a wide range of metallicity.  Therefore, the Phillips
relation is a convolution of whatever the dominant parameter is that
controls supernova light curves (e.g. central ignition density) and
the metallicity dependence of the local sample. Hence, it is
probably not surprising that the local sample does not show a
clear, systematic metallicity-dependent deviation from the mean Phillips
relation (Gallagher et al.\ 2005).

However, this does not imply that there should be no metallicity
effect at an earlier epoch. As long as there is more than one
parameter controlling SN Ia light curves, and as long as the supernova
sample changes with redshift, systematic deviations from the nearby
relation are expected. This is illustrated in Figure~2, which
schematically shows how sampling effects can affect the correction
function that needs to be applied. Note, in particular, that it is
possible to have a similar distribution of the main observable
parameter (e.g. $\Delta M_{15}$) for the local and the high-$z$
sample, but still have a systematic bias in the high-$z$
sample\footnote{The consequences of sampling different populations is
less straightforward when the stretch method instead of the $\Delta
M_{15}$ method is applied, as both local and high-$z$ supernovae are
used in the calibration (see, e.g., Astier et al.\ 2006).}.

In this paper, we will not attempt to correct for an intrinsic
metallicity dependence in the Phillips relation and assume instead
that the Phillips relation is just a function of the dominant
first parameter, independent of metallicity. We suspect that
this assumption will somewhat exaggerate any evolutionary
effects due to metallicity.

\subsection*{\em Mimicking $\Lambda$-dominated cosmologies}

\begin{figure}
\centerline{ \psfig{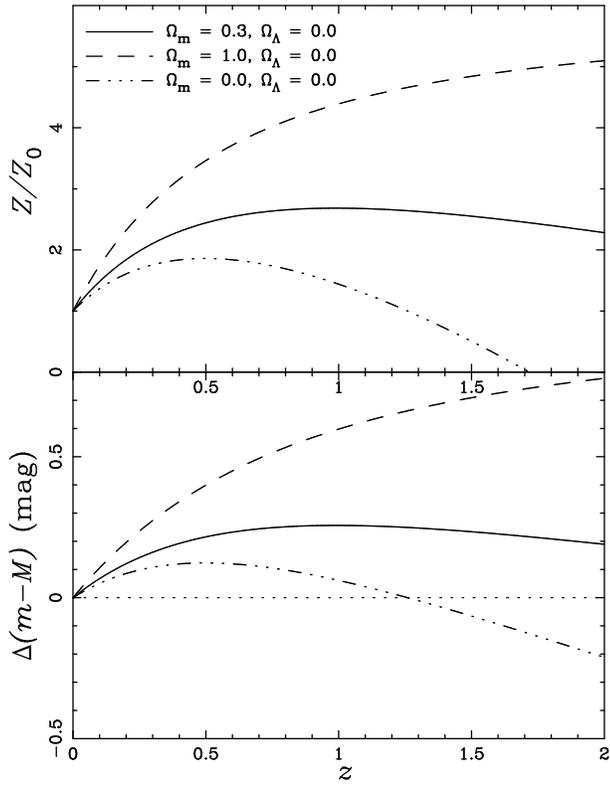} }
\caption{Mimicking $\Lambda$-dominated cosmologies. {\em Top panel:}
the evolution of the mean metallicity ($Z$) of the supernova sample as
a function of redshift ($z$) required to mimic a $(\Omega_{\rm
m}=0.3;\,\,\Omega_{\Lambda}=0.7)$ Universe in a $\Lambda$-free
Universe with $\Omega_{\rm m} = 0.0$ (empty Universe), 0.3 (open
Universe) and 1 (flat Universe), as indicated.  $\Omega_{\rm m}$ and
$\Omega_{\Lambda}$ are the dimensionless densities of matter and
``dark energy'', respectively, $Z_0=0.02$, and the present-day Hubble
parameter is assumed to be $70\,$km/s/Mpc. {\em Bottom panel:} the
effect of metallicity on the estimate of the distance modulus for the
three cases in the top panel, based on equation~(10) with
$\alpha=0.111$ and $Z_{\rm ref}= Z_0$.}

\end{figure}

In Figure~3, we illustrate the metallicity evolution that would mimic
a flat, $\Lambda$-dominated Universe with $\Omega_{\rm m} =0.3$ and
$\Omega_{\Lambda}=0.7$ in different $\Lambda$-free cosmologies as
indicated (here $\Omega_{\rm m}$ is the present matter density as a
fraction of the critical density and $\Omega_{\Lambda}\equiv \Lambda/3
H_0^2$, where $\Lambda$ is the cosmological constant and $H_0$ is the
present-day Hubble parameter, assumed to be $70\,$km/s/Mpc). These
curves were obtained by calculating the difference in distance modulus
as a function of redshift $z$ between the reference
$\Lambda$-dominated cosmology and the respective $\Lambda$-free
cosmologies (bottom panel of Fig.~3) and then solving equation~(10)
for $Z$, where we adopted $\alpha=0.111$ and assumed that the
present-day reference metallicity is $Z_0$ (i.e. $Z_{\rm
ref}=Z_0$). For example, if in an open Universe with $\Omega_{\rm
m}=0.3$, the mean metallicity of the supernova sample increased from
$Z_0$ at $z=0$ to about $2.5\,Z_0$ at $z=1$, as shown in the top
panel, this would exactly mimic a $\Lambda$-dominated Universe with
$\Omega_{\rm} = 0.3$ and $\Omega_{\Lambda}=0.7$\footnote{If the
reference metallicity were $0.5\,Z_0$, the metallicity would have to
increase to about $2\,Z_0$, since to lowest order it is change in
$\Delta Z$ that matters (see eq.~11).}.

To relate this metallicity dependence of the supernova sample to the
metallicity of stellar populations as a function of redshift, one also
needs to consider that there may be a significant time delay between
the formation of the supernova progenitor system and the actual
supernova. While the typical time delay in one of the most popular
single-degenerate models is only $\sim 1\,$Gyr, the fact that SNe Ia
also occur in old elliptical galaxies suggests that there should also
be a population of SNe Ia with relatively long time delays (e.g.\
Cappellaro \& Turatto 1988; Branch \& van den Bergh 1993). This could
be either due to the tail of the time-delay distribution if there is a
single dominant progenitor channel (e.g. in the double-degenerate
channel) or due to a second progenitor channel (e.g. Hachisu et al.\
1996; for a detailed discussion, see F\"orster et al.\ 2006).

\begin{figure}
\centerline{ \psfig{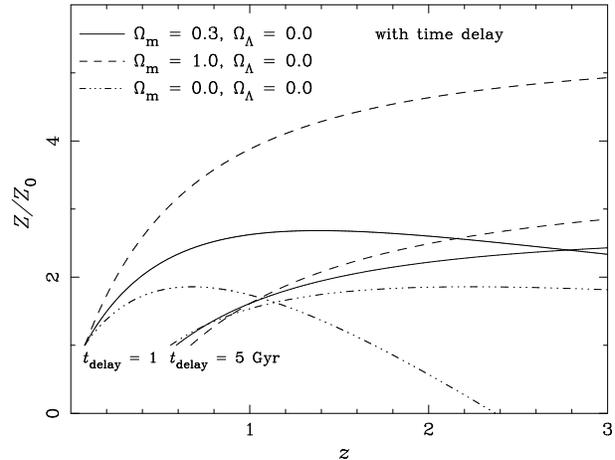} }
\caption{The mean metallicity evolution of the underlying star-forming
population as a function of redshift, mimicking a $\Lambda$-dominated
Universe for the cases shown in Figure 2. The two families of curves
assume a time delay of 1 and 5 Gyr between the star-formation epoch
and the supernova, as indicated.}
\end{figure}

In Figure~4, we show the evolution of the metallicity of the star-forming 
population with redshift that produces the metallicity dependence in
Figure~3, assuming a short time delay of 1 Gyr and a long time delay of 5 
Gyr\footnote{Note that for a given time delay, no SN Ia would occur beyond
a particular redshift $z_{\rm max}$ at which the time that has passed 
since the first star formation is equal to the time delay.}.

\subsection*{\em An open, $\Lambda$-free Universe with 
\boldmath{$\Omega_{\rm m}=0.3$}?}

As is clear from Figs 3 and 4, in order for the metallicity evolution
to mimic a $\Lambda$-dominated Universe, the metallicity has to
increase with redshift, since this makes observed supernovae
intrinsically fainter than is assumed.  This is at first sight rather
counter-intuitive. After all, the global metallicity in baryonic
matter (stars and gas) has to increase as the Universe
evolves. However, this need not apply for the mean metallicity in
star-forming regions. In the now widely accepted picture of galaxy
downsizing (e.g. Treu et al.\ 2005), the most massive galaxies form
first and fast and have largely finished star formation at a redshift
larger than 1 (e.g. Cowie 1996; Kauffmann et al.\ 2003). But the most
massive galaxies are also the galaxies with the largest
metallicities (e.g. Kobulnicky \& Kewley 2004). Thus, it is possible
that, as the Universe evolves and star-formation predominantly occurs
in galaxies of lower mass (with lower metallicities), the metallicity
in the star-forming regions decreases while the global metallicity
increases. Indeed, Panter, Heavens \& Jimenez (2003) (also see Sheth
et al.\ 2006) have claimed that the metallicity in star-forming
regions has declined for the last 6\,Gyr, based on their
reconstruction of the star-formation and metallicity history of the
galaxies in the Sloan Digital Sky Survey. However, this method
is not without its uncertainties (e.g., Mathis, Charlot \& Brinchmann 2006).

\begin{table}
\caption{The effects of time delays.}
\begin{tabular}{lcccc}
\hline\hline
\noalign{\vspace{2pt}}
cosmology&$t_0(z=0.8)$&\multicolumn{3}{c}{$z(t_0+\Delta t)$}\\
&(Gyr)&$\Delta t = 1\,$Gyr& $3\,$Gyr&$5\,$Gyr\\
\hline
\noalign{\vspace{2pt}}
\multicolumn{5}{l}{{\em no cosmological constant} ($\Omega_\Lambda =0$)}\\
$\Omega_{\rm m}=$\\
%\noalign{\vspace{2pt}}
\phantom{$\Omega_{\rm m} =$}0.0&6.2&1.1&1.9&4.1\\
\phantom{$\Omega_{\rm m} =$}0.3&5.9&1.1&2.3&$> 5$\\
\phantom{$\Omega_{\rm m} =$}1.0&5.5&1.2&3.9&--\\
\noalign{\vspace{2pt}}
\multicolumn{5}{l}{\em variation of the equation of state for flat cosmologies:}\\
($\Omega_{\rm m}$, $w_0$, $w_a$)\\
\noalign{\vspace{2pt}}
(0.30, $-1.0$, \ 0)&6.8& 1.0&1.8& 3.7\\
(0.35, $-1.0$, \ 0)&6.7& 1.0&1.9& 4.5\\
(0.20, $-1.0$, \ 0)&7.2& 1.0&1.6& 2.7\\
(0.30, $-1.4$, \ 0)&7.3& 1.0&1.7& 3.4\\
(0.30, $-0.8$, \ 0)&6.6& 1.0&1.8& 4.2\\
(0.30, $-1.0$, \ 1)&6.6& 1.0&1.9& 4.9\\
(0.30, $-1.0$,$-2$)&7.2& 1.0&1.7& 3.3\\
\noalign{\vspace{2pt}}
\hline
\end{tabular}\\
\parbox{8cm}{
\noindent{Note. ---} For the given cosmological parameters,
$t_0$ is the lookback time at a redshift $z=0.8$, 
$z(t_0+\Delta t)$ gives the redshift at lookback times $t_0$ plus
time delays of 1, 3 and 5\,Gyr, as indicated.}
\end{table}
 
\begin{figure*}
\centerline{ \psfig{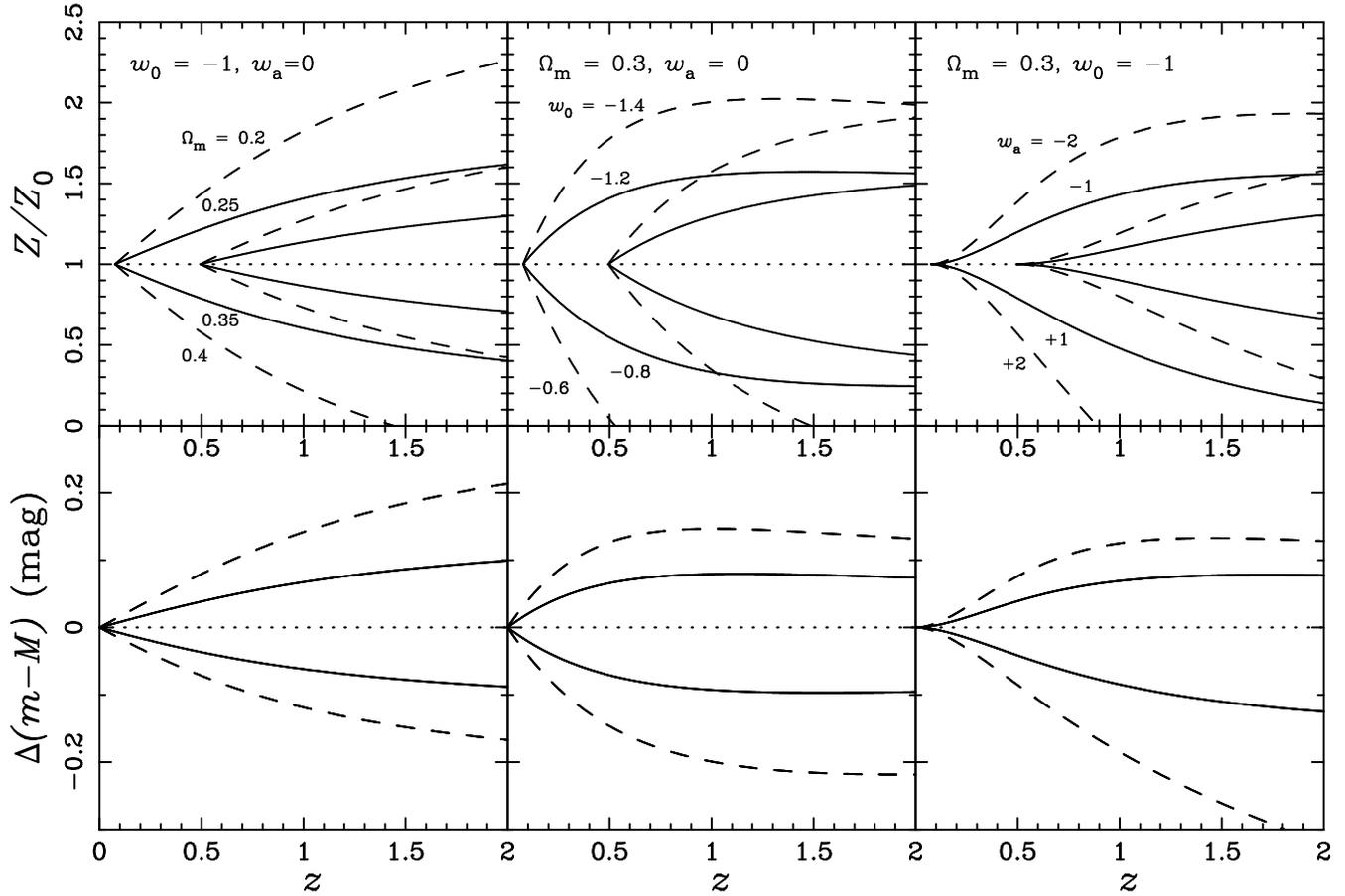} }
\caption{The effects of metallicity evolution on the determination of
cosmological parameters in a flat, $\Lambda$-dominated Universe.  The
underlying (true) cosmology is assumed to be a flat Universe with
$\Omega_ {\rm m}=0.3$, $\Omega_{\Lambda}=0.7$ and no evolution of the
equation of state (i.e.\ with $w_0=-1$ and $w_a=0$, where the equation
of state is parametrized as $w\equiv P/(\rho\,c^2) = w_0+w_a\,(1-a)$,
and $a=1/(1+z)$ is the dimensionless scale factor of the Universe).
The top panels show the mean metallicity evolution in the star-forming
component that would mimic the indicated deviation from the reference
model, while the bottom panels give the effect on the inferred
distance modulus (the two families of curves in the top panels are for
time delays of 1\,Gyr [left] and 5\,Gyr [right], respectively). For
each of the three sets of models, two of the three cosmological
parameters $\Omega_{\rm m}$, $w_0$ and $w_a$ are kept fixed, while the
third is varied as indicated ($\Omega_{\Lambda}=1 - \Omega_{\rm m}$ in
all cases, and $Z_0=0.02$).}
\end{figure*}

More direct evidence that the metallicity in star-forming regions may
have been higher in the past than at the present time comes from the
archaeological reconstruction of the metallicity history in early-type
galaxies (Thomas et al.\ 2005). According to Thomas et al.\ (2005),
the most massive early-type galaxies have a metallicity of 2 to 3
times solar and formed most of their stars between a redshift of 1.5
and 2 in low-density environments and a redshift up to 5 in the
densest environments (see also di Serego Alighieri, Lanzoni \& J\o
rgensen 2006).  Thomas et al.\ (2005) argue that the increase in
metallicity is mainly due to an increase in $\alpha$ elements rather
than an increase in iron-peak elements. However, since the oxygen
abundance is the most important abundance in determining the neutron
excess, this is directly applicable to our analysis\footnote{We note
that the underlying evolutionary tracks generally used in studies of
this type tend to use scaled solar abundances, where iron-peak
elements contribute significantly to the opacity. The effect of using
$\alpha$-enhanced (Fe-deficient) tracks should be to somewhat increase
the inferred metallicity and effective oxygen abundance, i.e. further
enhance the metallicity effect on SN Ia light curves.}.

In order to mimic a $\Lambda$-dominated Universe in an open
$\Omega_{\rm m}=0.3$ Universe, this only requires an increase in the
metallicity from 1\,$Z_0$ to $\sim 2.5\,Z_0$, well within the maximum
plausible metallicity range in early-type galaxies. However, at a
redshift of $\sim 0.8$ where the SNe Ia with the largest metallicity
are required, the most massive galaxies have already completed most of
their star formation. Therefore understanding the time delay between
the formation epoch of the progenitors of SNe Ia and the actual
explosion becomes essential.  If there is a systematic time delay of a
few Gyr, the peak in the metallicity--redshift relation (the solid
curves in Figs 3 and 4) can easily be shifted into the redshift range
where massive early-type galaxies form most of their stars (also see
Table~1). We note that even though the local sample of SNe Ia shows a
strong correlation with star formation (e.g. Mannucci et al.\ 2005),
implying a relatively short time delay of $\la 1\,$ Gyr (see F\"orster
et al.\ 2006), this need not be the case for the high-$z$ sample where
a different population, even one with a longer time delay, could
dominate; this depends entirely on the details of the star-formation
and metallicity histories and their effects on the SN Ia rate.  We
conclude that considering the present uncertainties in the metallicity
evolution of star-forming regions, a systematic metallicity trend as
large as required for an $\Omega_{\rm m}=0.3$, open Universe cannot be
ruled out.

\subsection*{\em A closed, $\Lambda$-free Universe with 
\boldmath{$\Omega_{\rm m}=1.0$}?}

In contrast to the $\Omega_{\rm m}=0.3$ case, for a flat,
$\Lambda$-free Universe, the metallicity of the supernova sample has
to increase from $Z_0$ to about $4\,Z_0$ at $z=0.8$. This is a much
larger effect than seems plausible, at least for $\alpha=0.111$.  To
produce such a large effect, one would probably require the maximum
value for our estimated range of $\alpha$ (i.e. $\alpha\simeq0.165$)
{\em and\/} that the time delays are fine-tuned so as to maximise the
effect.  This would essentially require that all SNe Ia at a redshift
around $z=0.8$ occurred in the most massive early-type galaxies or at
least in the bulges of galaxies. This can almost certainly already be
ruled out from a comparison of local and high-z host galaxies (e.g.,
Strolger et al.\ 2004; Sullivan et al.\ 2006), which does not show such
a dramatic trend in host galaxy properties.  However, we caution that
there are still significant uncertainties in our estimate of the
metallicity evolution effect, due to uncertainties in the value of
$\alpha$, the effective metallicity and its calibration to be used in
high-$z$ galaxies and other cumulative metallicity effects.  We
estimate that this uncertainty is about a factor of 2 {\em in either
direction}.

\section{Measuring the equation of state of dark energy}

Independent of the SN Ia data, there is now ample observational
evidence for a globally flat Universe from measurements of
anisotropies in the microwave background (WMAP; Spergel et al.\ 2003;
2006) which, when combined with other observational constraints
(e.g. from galaxy clustering and weak lensing), provides strong
independent evidence for a flat Universe dominated by dark energy (see
the discussion and references in Spergel et al.\ 2006). It has been
argued that one of the main challenges in modern cosmology is to
understand the nature of this dark energy.  Future surveys, now under
consideration, will attempt to constrain the physical nature of this
new form of energy by measuring its cosmological evolution. It is the
purpose of this section to quantitatively estimate the systematic
uncertainties metallicity evolution is likely to introduce when
measuring the equation of dark energy in future supernova surveys such
as SNAP (SuperNova/Acceleration Probe; Aldering et al.\ 2002;
also see Riess \& Livio 2006 for a complementary study of evolutionary
effects).

To describe the time evolution of the equation of state of dark energy,
we use the parametrization suggested by Linder (2003),
\begin{equation}
w(a) \equiv {P\over \rho c^2} = w_0 + w_a\,(1-a),
\end{equation}
where $a=1/(1+z)$ is the dimensionless scale factor of the Universe, the
parameter $w_0$ gives the present value of the equation
of state, and $w_a$ parametrizes its evolutionary history
(for a cosmological constant, $w_0=-1$ and $w_a=0$).
For a flat Universe, the Friedmann equation then becomes
\vbox{
\[
H^2\equiv\left({\dot{a}\over a}\right)^2 = H_0^2\,\left[
\Omega_{\rm m}\,a^{-3}\right.\phantom{nothing}\nonumber
\]
\nobreak
\begin{equation}
\phantom{left and noth}
+ \left.(1-\Omega_{\rm m})\,a^{-3\,(1+w_0+w_a)}\,
e^{-3w_a\,(1-a)}\right],
\end{equation}}
where $H_0$ is the present-day Hubble parameter (again assumed to be
70\,km/s/Mpc). In this parametrization, the evolution is completely
determined by three parameters $\Omega_{\rm m}$, $w_0$ and $w_a$
(keeping $H_0$ constant). In order to estimate the systematic
uncertainties due to metallicity evolution, we adopt as a reference
model a cosmology with a cosmological constant with $\Omega_{\rm
m}=0.3$, $w_0=-1$ and $w_a=0$, and then calculate which metallicity
evolution would mimic a deviation from this reference
model. Specifically, we keep two of three cosmological parameters
fixed and vary the third. The results of this exercise are presented
in Figure~5, which in the top panels shows the necessary metallicity
evolution in the star-forming material (for supernova time delays of 1
and 5\,Gyr) as a function of redshift and in the bottom panels the
implied effect on the estimated distance modulus.

In order to estimate the systematic errors introduced by metallicity
evolution, one in principle needs an estimate for the range of
metallicity variation that is ``reasonable''. Following the
discussions in the previous section, we make the simple and
conservative assumption that a variation of $Z$ from a few tenth $Z_0$
to $\sim 2\,Z_0$  could be plausible. We can then directly read off
rough estimates for the systematic errors for each parameter from
Figure~5. We obtain $\Delta \Omega_{\rm m} = -0.1/+0.05$, $\Delta w_0
= -0.4/+0.2$, and $\Delta w_a = -2/+1$. With some prior assumptions
about the metallicity evolution, one would obviously be able
to constrain these errors further.

\begin{figure}
\centerline{ \psfig{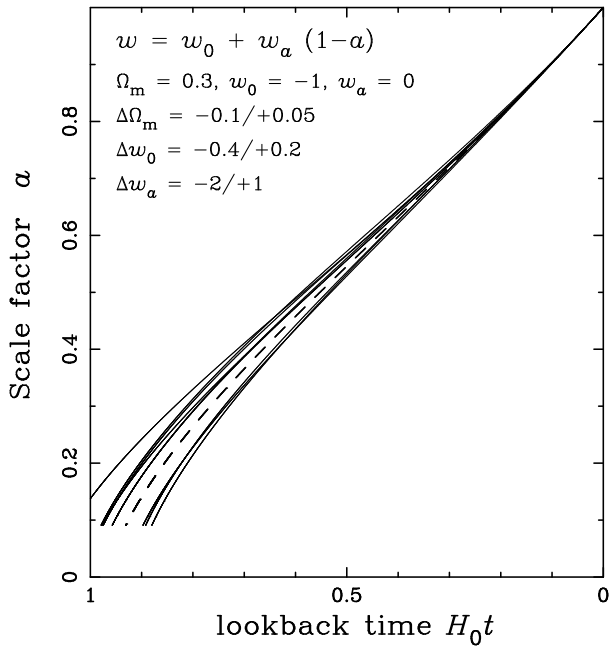} }
\caption{The evolution of the scale factor as a function of lookback
time (in units of $1/H_0$) for a variety of equations of state that
might be mimicked by metallicity evolution (see Fig.~5). The true
underlying cosmology is a flat $\Lambda$-dominated Universe with
$\Omega_{\rm m}=0.3$, $w_0=-1$ and $w_a=0$ (dashed curve). For the
individual solid curves, one of the three parameters was varied within
a range that could plausibly be caused by metallicity evolution (as
indicated).}
\end{figure}

To illustrate the uncertainty this could introduce in determining the
nature of dark energy through its time evolution, we plot in Figure~6
the scale factor as a function of lookback time (similar to Fig.~1 in
Linder 2003).  Here the dashed curve represents the evolution of the
scale factor for the assumed underlying cosmology (a flat Universe
with a cosmological constant), while the solid curves show the
evolution of $a$ for cosmologies that could plausibly be mimicked by
metallicity evolution. Comparing this figure to Figure~1 in Linder
(2003), we note that there is a large overlap of these curves with the
curves produced by different physical models of dark energy. This
implies that it will be difficult to distinguish these different
models, unless metallicity evolution effects are taken into account
and are corrected for.

Our main conclusion therefore is that the possible systematic
uncertainties in $\Omega_{\rm m}$ and $w_0$ caused by metallicity
evolution could be quite significant, but may be manageable (in
particular, if further constraints on the metallicity evolution can be
employed). In contrast, the uncertainty in $w_a$ seems uncomfortably
large, implying that it may be very difficult to differentiate time
evolution in the equation of state from time evolution in $Z$ using SN
Ia data alone.

\section{Discussion and Conclusions}

Our main conclusion is that metallicity should affect the SN Ia
calibration method; this relies only on reasonably straightforward and
well understood nuclear physics. As long as there is {\em any}
significant cosmological evolution in the metallicity, which
{\em a priori} is to be expected, this has to affect the
determination of cosmological parameters {\em at some level}.  The
main remaining uncertainty is the level at which this becomes
important.

In our estimates of systematic errors, we adopted $\alpha = 0.111$,
roughly twice the estimate of Timmes et al.\ (2003), though $\alpha$
could possibly as large as 0.165. It still remains to be shown
theoretically that the metallicity effect is as large as these
estimates suggest. This will require detailed stellar evolution
calculations of the pre-explosion phase that model the convective Urca
process realistically (Lesaffre et al.\ 2005 and references therein)
combined with up-to-date explosion modelling (e.g.\ R\"opke et al.\
2005), work that is presently in progress (F. F\"orster, et al.). In
general, since the magnitude corrections scale linearly with $\alpha$,
at least to lowest order (see eq.~11), all estimates of systematic
errors in this paper can be rescaled with whatever the appropriate
value for $\alpha$ will turn out to be.

It would be particularly useful to be able to constrain the value of
$\alpha$ observationally. This is quite a challenging task even for
the local SN Ia sample, since it is clear that metallicity cannot be
the primary parameter controlling SN Ia light curves (Hamuy et al.\
2000; Gallagher et al.\ 2005) and without a better understanding of
the dominant parameter and how it affects the basic light curve
correction method, it will be difficult to isolate the metallicity
contribution. On the other hand, since the predicted
metallicity-dependent correction factor is, to lowest order, a linear
function of metallicity, with a sufficiently large sample of SNe Ia
with well determined host galaxy metallicities, one could include a
linear metallicity term in the Phillips relation and attempt a
two-parameter calibration. This could in principle allow a measurement
of the magnitude of the metallicity effect (i.e. constrain $\alpha$)
and could then be used to correct for metallicity effects at high
$z$\footnote{Obtaining the metallicity of the SN Ia progenitor
directly is probably not feasible, and therefore one has to rely on a
statistical measure of the metallicity using the host galaxy
metallicity. This itself is not without problems, considering, e.g.,
that there may be a substantial time delay between the formation of
the SN progenitor and the explosion and that different metallicity
methods have to be employed for different types of galaxies, all with
their own calibration issues. In particular, it will generally be
difficult to obtain a reliable metallicity estimate for the
progenitors of SNe Ia in early-type galaxies, since the formation of
the supernova progenitor may be associated with a relatively recent
merger event, in which case its metallicity need not bear any relation
to the metallicity of the dominant old population (e.g. Nolan et al.\
2006).}.

More progress is likely to come from a systematic comparison of the
properties of nearby and high-z supernovae, one of the major
objectives of both the ESSENCE project\footnote{See
http://www.ctio.noao.edu/\~\,wsne/index.html.}  and the SuperNova
Legacy Survey (SNLS)\footnote{http://cfht.hawaii.edu/SNLS.}. First
results from the SNLS (Astier et al.\ 2006; Sullivan et al.\ 2006)
indicate that the nearby and high-z samples are similar, but that
there may be some differences in properties. While at present these
are only of marginal statistical significance, this may change as the
sample increases.

The evolution of galaxy properties, in particular the metallicity in
the star-forming regions, clearly plays an important role in the
theoretically predicted evolution of SN Ia properties.  There is a
general consensus that there has been a significant evolution of
galaxy properties since a redshift of 2\,--\,3: star formation first
occurred in the most massive galaxies and since then has moved
increasingly towards lower-mass galaxies.  Observationally, this is
reflected in star formation predominantly occurring in ultraluminous
infrared galaxies (ULIRGs) at a redshift $z\ga 2$, in luminous
infrared galaxies (LIRGs) at a redshift $z\simeq 1$ and starburst
galaxies in the local Universe (see, e.g., Fig~10 in
P\'erez-Gonz\'alez et al.\ 2005). If this galaxy trend is also
correlated with a trend in metallicity, it would exactly produce the
type of evolutionary effect that could affect the determination of
cosmological parameters.  A better understanding of the metallicities
in these different types of galaxies is therefore clearly needed.

In summary, metallicity effects should be expected and, at some level,
will affect the determination of cosmological
parameters. Uncertainties in the metallicity evolution in star-forming
galaxies and in the intrinsic properties of SN Ia progenitors (in
particular, their time delays) will limit the precision with which the
equation of state can be measured; it will be particularly difficult
to distinguish an evolution of the equation of state from evolution in
SN Ia properties. As the statistical errors in SN Ia surveys and
complementary surveys (CMB, galaxy clustering, weak lensing) decrease,
this may first show up in statistical discrepancies in the
determination of cosmological parameters using different
methodologies.

Our estimates show that metallicity evolution could possibly be large
enough to mimic a $\Lambda$-dominated Universe in an open Universe,
but is unlikely to be large enough to mimic a preferred flat
Universe. Nevertheless, considering the fundamental shift in our view
of the physical world adoption of a cosmology dominated by dark energy
would bring about, we emphasize the importance of taking such
systematic effects into account. One can hope that with a better
understanding of such effects, one will be able to correct for these
and ultimately obtain a reliable calibration of cosmological
parameters, where systematic uncertainties are minimal.

\section*{Acknowledgements}
We thank numerous colleagues for useful discussions and encouragement;
these include James Binney, Justin T. Bronder, Stephen Justham,
Claudia Maraston, Saul Rappaport, Kevin Schawinski, Daniel Thomas,
J. Craig Wheeler and Simon M. White. F.F. was supported by a
Fundaci\'on Andes -- PPARC Gemini studentship, C.W. by a PPARC
Advanced Fellowship.  This work was also supported in part by a
European Research \& Training Network on Type Ia Supernovae
(HPRN-CT-20002-00303).

%%%%%%%%%%%%%%%%%%%%%%%%%%%%%%%%%%%%%%%%%%%%%%%%%%%%%%%%%%%%%%%%%%%%%%%%%%%%%%%

\end{document}